\documentclass[a4paper,11pt]{article}
\usepackage{latexsym}	
\usepackage{amsmath}
\usepackage{amsthm}
\usepackage{graphicx}
\usepackage[dvipsnames]{xcolor}
\def\ba{\begin{eqnarray}}
\def\ea{\end{eqnarray}}

\def\ba{\begin{eqnarray}}
\def\ea{\end{eqnarray}}

\def\lb{\label}
\def\be{\begin{equation}}
\def\ee{\end{equation}}

\theoremstyle{plain}

\begin{document}
\title{Peculiarities for domain walls in Taub coordinates}
\author{Leandro Lanosa\thanks{
lanosalf@gmail.com and lflanosa@dm.uba.ar} \and Osvaldo Pablo Santillán\thanks{firenzecita@hotmail.com and osantil@dm.uba.ar.}}
\date{%
    Instituto de Matem\'atica Luis Santal\'o (IMAS), UBA CONICET, Buenos Aires, Argentina\\[2ex]%
    }
\maketitle
\begin{abstract}
In the present letter the infinite domain wall geometry in GR \cite{vilenkin1}-\cite{ipser} is reconsidered in Taub coordinates \cite{taub}. The use of these coordinates makes explicit that the regions between the horizons and the wall and the outer ones are flat.
By use of these coordinates, it is suggested that  points inside the horizon and outside never communicate each other. The wall is seen on the left and the right side as contracting and expanding portions of spheres and a plane singularity, which is the imprint the contracting and expanding domain wall. Particles of each region will never reach this imprint. In addition, at some point during the evolution of the system, four curious holes inside the space time appear, growing at the speed of light. This region is not parameterized by the standard Taub coordinates, and the boundary of this hole adsorbs all the particles that intersect it. The boundary of these holes
are composed by points which in the coordinates of \cite{vilenkin1}-\cite{ipser} are asymptotic, in the sense that they correspond to trajectories tending to infinite values of the time or space like coordinates, while the proper time elapsed for the travel is in fact finite. This is not paradoxical, as the coordinates \cite{vilenkin1}-\cite{ipser}  are not to be identified with
the true lengths or proper time on the space time.  The correct interpretation of the boundary is particularity relevant when studying scattering of quantum fields approaching the domain wall. A partial analysis about this issue is done in the last section.

 \end{abstract}

 \section{Introduction}
 There is no entire consensus about the role of domain walls in cosmology. Several scenarios assume their presence at the early universe \cite{domwal1}-\cite{domwall18}, and the characteristics
 of these topological objects are dependent on the underlying Quantum Field Theory in consideration.
 There are numerical studies that suggest that such objects, if arising in a first order phase transition, enter in the scaling regime and their correlation length becomes of the order
 of the Hubble radius. Such objects easily overcome the critical energy density of the universe \cite{doma1}-\cite{doma2} and may be in conflict with the experimental data of gravitational and CMB radiation. The problem is that this data constraints the wall tension $\mu$ to extremely low value \cite{vilenkinshe}. Some proposed alternatives assume the breaking of the symmetry leading to the domain wall formation, which implies that the global minima will prevail after a while. An example are axion domain walls \cite{domwall14}, see also \cite{zhit} and references therein. Another possibility is to assume that the initial conditions are biased \cite{domwall15}, a  possibility that softnesses the problem.
Furthermore, there are scenarios with spontaneous nucleation of domain walls that may avoid this problem \cite{vilenkoso1}-\cite{vilenkoso2}. These scenarios can potentially accommodate 
primordial black hole and wormholes.  

The purpose of the present work however,  is not to address in detail the cosmological problems related to domain walls, although some words are going to be said. Instead, it is focused on some global
features of the space time of a simple infinite domain wall solution \cite{vilenkin1}-\cite{ipser}. In those original references, the local form of the solution was found and was extended 
to a region with infinite volume in spatial coordinates. The solution is composed by two regions, one on the left and one on the right of the domain wall, each containing an event horizon at some
critical distance from the wall. Each of the four resulting regions are not only Ricci flat, but flat as well, as already pointed out in \cite{vilenkin2} by use of some results of Taub \cite{taub}. In the present work it is suggested that the use of those Cartesian coordinates reveals some global features that are perhaps not evident  with the coordinates employed in \cite{vilenkin2}. In particular,  the particles in the inner and outer region do not communicate since 
no particle can cross the horizon. The situation is therefore different  than  black holes.
 In addition, some holes growing at speed of light appear in the space time. 
These holes will be described in the text, and it is suggested that the adsorb all the particles which collide them.

The distinction between true asymptotic regions from apparent ones is important, specially when studying the dynamics of scalar, vector and even graviton scattering
in the geometry, since this changes radically the boundary conditions imposed for tackling those problems. A partial analysis about this subtle point is presented here.
 
The present work is organized as follows. In section 2 the use of Taub coordinates is clarified and the asymptotic and apparently asymptotic regions are differentiated.
The mentioned holes in the geometry, which correspond to a false asymptotic region, are described in detail in this text. In section 3 the geodesics and the behavior of scalar fields in the geometry is partially analyzed, and the possibility
for the geodesics to be non smooth and the presence of conjugate points is pointed out. Section 4 contain the discussion of the results and their possible applications.

\section{The local properties of the space time of a vacuum domain wall}

\subsection{The metric and its flat regions}
Consider an infinite vacuum domain wall in GR, that is, the solution of the Einstein equations for matter distributed along the plane $yz$. Its energy momentum tensor is independent
of the $x$ coordinate and is locally given by  $T_{\mu\nu}=\mu (1,0,1,1)$ \cite{vilenkin1}. The corresponding gravitational field was found in \cite{vilenkin2} and it is given by 
\be\lb{domwall}
g=e^{-\mu |x|} (-dt^2+dx^2)+e^{\mu(t-|x|)}(dy^2+dz^2).
\ee
Here $\mu$ is the surface mass density on the object, which is located at $x=0$. The Newtonian limit shows that the gravitational field of the wall is indeed repulsive \cite{vilenkin2}.
The fact that the metric decays exponentially 
at infinite is non physical, and it is showing that these coordinates do not cover the whole space. However, coordinates that go beyond the region $x\to \pm \infty$ can be found separately for both the left
$x<0$ or the right hand $x>0$. As it will be shown below, following the original references, the metric is flat in these regions.

Consider first the region on the right of the wall  $x>0$. A new coordinate $\Xi$ can be defined through the integration of $d\Xi=e^{-\frac{\mu x}{2}}dx$. By choosing the integration constant in such a way that the wall location $x=0$
corresponds to $\Xi=0$, the coordinate becomes
$$
\Xi=\frac{2}{\mu}(1-e^{-\frac{\mu x}{2}}).
$$
This new coordinate $\Xi$ takes values in the interval $[0,2/\mu)$ when $x$ takes all the positive real values.
By further extending the range of $\Xi$ to the interval $[0,\infty)$, the metric becomes
\begin{equation}\label{metricrightinner}
g=-\bigg(1-\frac{\mu \Xi}{2}\bigg)^2 dt^2+d\Xi^2+e^{\mu t}\bigg(1-\frac{\mu \Xi}{2}\bigg)^2(dy^2+dz^2),
\end{equation}
This metric extends beyond the original region $x\to \pm \infty$, and it is an extension of (\ref{domwall}) to a larger region.
As for several black hole solutions, there is a horizon  at $\left|\Xi\right|=2/\mu$. This horizon corresponds to the apparent asymptotic region to $x\to \pm \infty$, which is in fact at finite
distance from the wall at $x=0$.

An important property is that in both regions inside or outside the horizon, there is a coordinate system converting the metric into the flat metric. 
The presence of these coordinates was noticed in \cite{vilenkin2} based on an early work \cite{taub}. The statement
is correct, however, there is some typos in some formulas of \cite{taub} and it is convenient to make an independent derivation here.

For future reference, the regions are:
\begin{align}
	&\text{Left outer region:} &\quad &\Xi<-2/\mu, \nonumber\\
	&\text{Left inner region:} &\quad -2/\mu<&\Xi<0, \nonumber\\
	&\text{Right inner region:} &\quad 0<&\Xi<2/\mu, \nonumber\\
	&\text{Right outer region:} &\quad 2/\mu<&\Xi. \nonumber
\end{align}

For each region, there is a set of coordinate transformations that leads to a flat metric in that region, but not in the others. The transformations can be found as follows. First, rewrite the last distance element as:
$$
g=e^{\mu t}\bigg(1-\frac{\mu \Xi}{2}\bigg)^2 \bigg[-e^{-\mu t}dt^2+e^{-\mu t} \frac{d\Xi^2}{\bigg(1-\frac{\mu \Xi}{2}\bigg)^2}+dy^2+dz^2\bigg].
$$

Consider first the \textit{right inner region} $0<\Xi<2/\mu$. Choose the new coordinates
\begin{equation}\label{phi,psi,innerright}
\Phi_i=\frac{2}{\mu} e^{-\frac{\mu t}{2}}, \qquad \Psi_i=\log\bigg(1-\frac{\mu \Xi}{2}\bigg),
\end{equation}
and write the metric, up to a constant factor, as
$$
g=\frac{e^{2\Psi}}{\Phi^2}\bigg[-d\Phi^2+\Phi^2 d\Psi^2+dy^2+dz^2\bigg].
$$
The hyperbolic parametrization
\be\lb{hiperbolic}
\tau=-\Phi_i \cosh\Psi_i,\qquad \chi_i=\Phi_i \sinh \Psi_i,
\ee
converts the last expression into
$$
g=\frac{1}{(\chi+\tau)^2}\bigg[-d\tau^2+d\chi^2+dy^2+dz^2\bigg].
$$
The further choice $\alpha=\chi-\tau$ and $\beta=\chi+\tau$ brings the last metric to the form 
\be\lb{cf}
g=\frac{1}{\beta^2}[d\alpha d\beta+dy^2+dz^2].
\ee
The transformation 
\be\lb{transf2}
 \alpha=-V-\frac{Y^2+Z^2}{U},\qquad \beta=\frac{1}{U}, \qquad y=\frac{Y}{U}, \qquad z=\frac{Z}{U},
\ee
with inverse
\be\lb{transf}
U=\frac{1}{\beta}, \qquad V=-\alpha-\frac{y^2+z^2}{\beta}, \qquad Y=\frac{y}{\beta}, \qquad Z=\frac{z}{\beta},
\ee
converts the last metric into 
\be\lb{flat}
g=dU dV+dY^2+dZ^2,\qquad U=X-T, \qquad V=X+T.
\ee
Thus the metric is flat in this region, as stated. The coordinates $U$ and $V$ are the standard retarded and advanced null coordinates for the flat Minkowski metric.

In the following, a suffix $i$ will be included in the coordinates denomination, in order to emphasize that they corresponds to the inner region. The explicit functional form for $\alpha_i$ and $\beta_i$ is given by
\be\lb{tata}
\alpha_i=\Phi_i e^{\Psi_i},\qquad \beta_i=-\Phi_i e^{-\Psi_i}.
\ee
By expressing $\Phi_i$ and $\Psi_i$ in terms of $\Xi$ and $t$ by the formulas given above, the last quantities can be written as follows
\be\lb{tata3}
\alpha_i=\frac{2}{\mu}e^{-\frac{\mu t}{2}}\bigg(1-\frac{\mu \Xi}{2}\bigg),\qquad \beta_i=-\frac{2}{\mu\bigg(1-\frac{\mu \Xi}{2}\bigg)}e^{-\frac{\mu t}{2}}.
\ee
Furthermore, from  (\ref{transf}) and (\ref{tata3}) it follows that
$$
U_i=-\frac{\mu}{2}\bigg(1-\frac{\mu \Xi}{2}\bigg)e^{\frac{\mu t}{2} }, \qquad V_i=-\frac{2}{\mu}\bigg(1-\frac{\mu \Xi}{2}\bigg)e^{-\frac{\mu t}{2} }+\frac{\mu(y^2+z^2)}{2}\bigg(1-\frac{\mu \Xi}{2}\bigg)e^{\frac{\mu t}{2} }, 
$$
\be\lb{transfw}
Y_i=-\frac{\mu y}{2}\bigg(1-\frac{\mu \Xi}{2}\bigg)e^{\frac{\mu t}{2} }, \qquad Z_i=-\frac{\mu z}{2}\bigg(1-\frac{\mu \Xi}{2}\bigg)e^{\frac{\mu t}{2} }.
\ee
Since $T=V-U$, it is seen from the last expression that $T\to \infty$ when $t\to \infty$ and vice versa, thus $T$ and $t$ have the same orientation.
The next task is to parameterize the boundaries in these coordinates. From the definition $\Psi_i=\log(1-\frac{\mu\Xi}{2})$ given above, it is clear that at the wall location $\Xi=0$ the coordinates
$\alpha_i$ and $\beta_i$ in (\ref{tata3}) satisfy  the simple relation
$$
\alpha_i=-\beta_i.
$$
This condition together with  (\ref{transf}) shows that, if $U_i\neq 0$, the wall in the flat coordinates
is composed by a set points satisfying
$$
X_i^2+Y_i^2+Z_i^2=1+T_i^2.
$$
This represents a sphere of variable radius in time, which contracts and expands to an infinite volume. In four dimensions, it is an hyperboloid. The question
is now to understand if the corresponding region is inside or outside this sphere.
It will be shown below that the points between the wall and the horizon are inside this sphere, since the origin will belong to the region. However, the region does not contain all the points of this sphere.
This can be seen as follows. From their definitions (\ref{tata}) it follows that $\alpha_i> 0$ and $\beta_i<0$.\footnote{Remember that these are defined at the inner right region $0<\Xi<2/\mu$ by $\Phi$ and $\Psi$ in eq. (\ref{phi,psi,innerright}), and that in any other region, i.e. right outer region or left inner and outer regions, $\Phi$ and $\Psi$ need to be redefined, and that cases must be considered separately.} From this fact, together with (\ref{transf}), it is concluded that
$$
X_i\leq T_i, \qquad T_i^2\leq X_i^2+Y_i^2+Z_i^2.
$$
The border is the region $X_i=T_i$ or, equivalently, $U_i=0$, and this is reached only asymptotically.
The first of the last inequalities implies that $U_i<0$, while the sphere $T_i^2= X_i^2+Y_i^2+Z_i^2$ corresponds to $\alpha_i=0$ with $\beta_i\neq 0, \infty$. Thus, the \textit{right inner region} are all the points bounded by two spheres $T_i^2\leq X_i^2+Y_i^2+Z_i^2\leq 1+T_i^2$, which satisfy the restriction $U_i< 0$. The bigger sphere, i.e. the wall, intersects the border $U_i=0$ in the form  of a circle $Y_i^2+Z_i^2=1$.
This circle moves at speed of light in the $X$ since $U=0$ means $X=T$. The small sphere only intersects $U=0$ at the points $Y=Z=0$. This can be seen at Figures 1 and 2.

In fact, union of the small sphere $X_i^2+Y_i^2+Z_i^2=T_i^2$ and the surface $U=0$ (depicted as a line in the figures) are both part of the horizon $\Xi=2/\mu$. This can be seen by taking the limit $\Xi \to 2/\mu \text{ and } t \to +\infty$ in such a way that $U_i=-\frac{\mu}{2}\bigg(1-\frac{\mu \Xi}{2}\bigg)e^{\frac{\mu t}{2} }\to c\neq 0$, which implies that $\alpha=0$ and  $\beta_i\neq 0, \infty$. As remarked above, this is exactly the small sphere. Secondly, by taking $\Xi \to 2/\mu$ values $U \to 0$ can be obtained in other several limits of $t$. Finally, it is appropriate to mention that part of the region $U=0$ is not in the horizon, as the values $U \to 0$ can be obtained in the limiting case $\Xi \neq 2/\mu \text{ and } t \to -\infty$. This corresponds to the asymptotic past of points inside the inner region. Both surfaces includes the origin $X=Y=Z=0$ of the flat space, which they intersects at $T=0$.



It is clear from the discussion above that the space $(T, X, Y, Z)$ defined by the flat metric (\ref{flat}) is an extension of the space $(t, \Xi, y, z)$
defined by the metric (\ref{metricrightinner}) and so, it is a  further extension of the original space $(t, x, y, z)$ with metric (\ref{domwall}).


It may be practical to characterize further  the small sphere. From its definition, it is seen that it contracts to the origin at light speed
and then expands with the same velocity. It intersects the plane $U_i=0$ only at the point $Y_i=Z_i=0$. However, as it is centered around the spatial origin
$X_i=Y_i=Z_i=0$ it is outside the inner region at negative times, as the condition $U_i<0$ implies that $X_i<T_i<0$. The spatial origin therefore
is not in the region for negative times and the sphere is outside. However, the outside region requires an independent analysis, which is to be done below and which will show that
this sphere is not present. Instead, when the time $T_i$ becomes positive, the spatial origin satisfies the condition $U_i<0$ and the sphere grows inside the region at light speed. Thus, the region apparently has a hole growing.
This is exemplified in Figures 1 and 2. A feature that may call the attention is  that the boundary of the hole can be reached from the points in the inner region  $T_i^2\leq X_i^2+Y_i^2+Z_i^2\leq 1+T_i^2$ at finite proper time, and therefore it appears
the question whether the particles which hit this boundary $T_i^2= X_i^2+Y_i^2+Z_i^2$ disappear,  or whether the region inside this sphere is filled and the original coordinates do not cover it. This will be discussed in the next subsection.


\begin{figure}
 \centering
\includegraphics[height=199.9722pt]{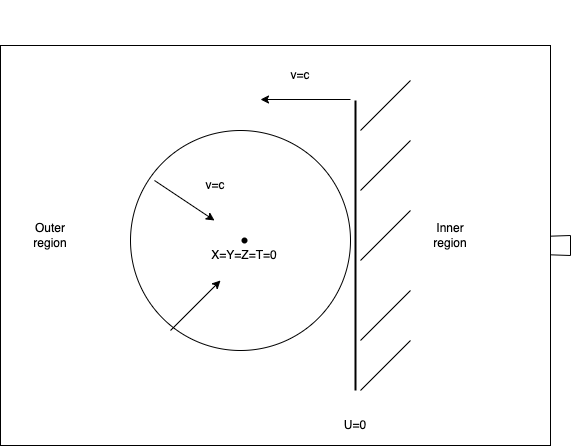}

\footnotesize
  \caption{The hole shrinking  outside the inner region for $T<0$. It is outside the region, and it does not exist, since this outside region has to be studied separately.}
\end{figure}



\begin{figure}
 \centering
\includegraphics[height=199.9722pt]{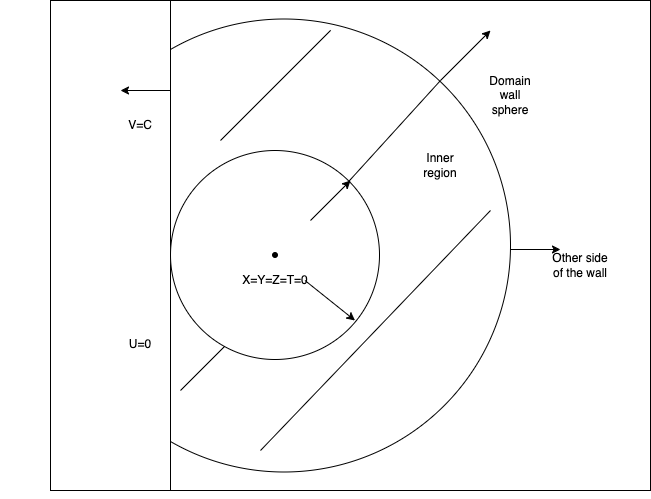}

\footnotesize
   \caption{The hole growing inside the left inner region for $T>0$. Since it grows at the light speed, the particles outside may collide the hole boundary, after that they are adsorbed at this boundary. The larger sphere represents the wall.
   The intersection of the small sphere with $U=0$ is the point $Y=Z=0$ while the sphere wall intersects it at the circle $Y^2+Z^2=1$. The right hand is exactly equal, but in order to see these figures the
   traveler has to change the shuttle.}
\end{figure}

An important consequence of the previous discussion is that a particle that crosses the horizon $X_i=Y_i=Z_i=T_i=0$ will never reach the wall, as it will be always
over the small sphere $T_i^2= X_i^2+Y_i^2+Z_i^2$. The converse of the previous statement is also true. Consider a light signal starting from the wall at $T_0\leq 0$ directly radially to the spatial origin. 
Its parametrization is $R=\sqrt{1+T_0^2}-(T-T_0)$, with a constant direction. This ray will reach the horizon  $X_i=Y_i=Z_i=T_i=0$ 
if $R=0$ at $T=0$. This condition is then $\sqrt{1+T_0^2}+T_0=0$ and it has no solutions, except $T_0\to -\infty$.

Let study now the \textit{left inner region}. The generic form of the metric valid on both sides of the wall is
$$
g=-\bigg(1-\frac{\mu | \Xi| }{2}\bigg)^2 dt^2+d\Xi^2+e^{-\mu t}\bigg(1-\frac{\mu | \Xi| }{2}\bigg)^2(dy^2+dz^2),
$$
and it is clear that it is invariant under $\Xi\to -\Xi$. The transformation making explicit the flatness of the metric on the left hand side  $0\leq-\Xi<2/\mu$ is now
$$
\widetilde{\Phi}_i=\frac{2}{\mu} e^{\frac{\mu t}{2}}, \qquad \widetilde{\Psi}_i=\log\bigg(1+\frac{\mu \Xi}{2}\bigg),
$$
The formulas then are analogous to the previous ones, with the replacement $\Xi\to -\Xi$, namely
\be\lb{tata33}
\widetilde{\alpha}_i=\frac{2}{\mu}e^{-\frac{\mu t}{2}}\bigg(1+\frac{\mu \Xi}{2}\bigg),\qquad \widetilde{\beta}_i=-\frac{2}{\mu\bigg(1+\frac{\mu \Xi}{2}\bigg)}e^{-\frac{\mu t}{2}}.
\ee
The region in the space time is given by
$$
\widetilde{T}_i^2\leq\widetilde{X}_i^2+\widetilde{Y}_i^2+\widetilde{Z}_i^2\leq 1+\widetilde{T}_i^2.
$$
Here the wide tilde denote the quantities on the left hand side.
In addition $\widetilde{U}_i\leq 0$ in this inner region, and the boundary is $\widetilde{U}_i=0$. Furthermore
$$
\widetilde{U}_i=-\frac{\mu}{2}\bigg(1+\frac{\mu \Xi}{2}\bigg)e^{\frac{\mu t}{2}}, \qquad \widetilde{V}_i=-\frac{2}{\mu}\bigg(1+\frac{\mu \Xi}{2}\bigg)e^{-\frac{\mu t}{2}}+\frac{\mu(y^2+z^2)}{2}\bigg(1+\frac{\mu \Xi}{2}\bigg)e^{\frac{\mu t}{2}}, 
$$
\be\lb{transfw2}
\widetilde{Y}_i=-\frac{\mu y}{2}\bigg(1+\frac{\mu \Xi}{2}\bigg)e^{\frac{\mu t}{2}}, \qquad \widetilde{Z}_i=-\frac{\mu z}{2}\bigg(1+\frac{\mu \Xi}{2}\bigg)e^{\frac{\mu t}{2}}.
\ee

It should be remarked that the coordinates on the left and right hand side takes the same set of values. To see this, note that for the same values of $y$, $z$ and $t$ the points on the right with $\Xi$
and the point of the left with $-\Xi$ have the same values of $U$, $V$, $Y$ and $Z$. That is the reason for employing the wide tilde for the quantities on the left. Both sides are identical, and for positive times an sphere growing at light speed centered at the origin appears. 

A feature that may be slightly curious is that, at positive times, both the observers on the left and on the right see the wall location as an expanding piece of sphere, but are both inside it.
The point is that the coordinates that make flat the left region does not do the job on the right.  In other words, an observer which sees a flat space on the left do not see a flat one on the right when crossing the wall, and has to change to a new shuttle to do it, i.e. a different set of coordinate transformations are needed to make the other region flat. At this new shuttle an expanding sphere is seen.

The asymptotic region $U=0$ moves at speed of light, and the points inside the region between the wall and horizon are those with $U\leq 0$. An observer located at any of these points never reaches this boundary at finite proper time, since
it is running away at speed of light. 

The next task is to study the \textit{right} and \textit{left outer regions}, and the coordinates that make flatness in each region are respectively:
$$ 
U_o=-\frac{\mu}{2}\bigg(\frac{\mu \Xi}{2}-1\bigg)e^{\frac{\mu t}{2}}, \qquad V_o=-\frac{2}{\mu}\bigg(\frac{\mu \Xi}{2}-1\bigg)e^{-\frac{\mu t}{2}}+\frac{\mu(y^2+z^2)}{2}\bigg(\frac{\mu \Xi}{2}-1\bigg)e^{\frac{\mu t}{2}}, 
$$
\be\lb{transfw3}
Y_o=-\frac{\mu y}{2}\bigg(\frac{\mu \Xi}{2}-1\bigg)e^{\frac{\mu t}{2}}, \qquad Z_o=-\frac{\mu z}{2}\bigg(\frac{\mu \Xi}{2}-1\bigg)e^{\frac{\mu t}{2}},
\ee
and
$$ 
\widetilde{U}_o=-\frac{\mu}{2}\bigg(\frac{\mu \Xi}{2}+1\bigg)e^{\frac{\mu t}{2}}, \qquad \widetilde{V}_o=-\frac{2}{\mu}\bigg(\frac{\mu \Xi}{2}+1\bigg)e^{-\frac{\mu t}{2}}+\frac{\mu(y^2+z^2)}{2}\bigg(\frac{\mu \Xi}{2}+1\bigg)e^{\frac{\mu t}{2}}, 
$$
\be\lb{transfw4}
\widetilde{Y}_o=-\frac{\mu y}{2}\bigg(\frac{\mu \Xi}{2}+1\bigg)e^{\frac{\mu t}{2}}, \qquad \widetilde{Z}_o=-\frac{\mu z}{2}\bigg(\frac{\mu \Xi}{2}+1\bigg)e^{\frac{\mu t}{2}}.
\ee
where the suffix $o$ means they corresponds to the outer regions, and the wide tilde coordinates are the ones corresponding to the left side of the wall.

Here the space time is given by the points with $\widetilde{U}_o\leq 0$ and 
$$
\widetilde{T}_o^2\leq\widetilde{X}_o^2+\widetilde{Y}_o^2+\widetilde{Z}_o^2,
$$
which follows from the fact that $\widetilde{\alpha}_o\leq 0$. Thus, an sphere expanding with at the speed of light  is also present in those two outer regions and the particles living in them
 do not reach the boundary $\widetilde{U}_o=0$. The inner and outer region are therefore not connected.
 The graphs in Figure 3 describes the geometry with their holes, adapted to the coordinates that makes the metric in the left inner region Minkowski.
 
 \begin{figure}\label{fig:all regions}
 \centering
   \includegraphics[height=277.9722pt]{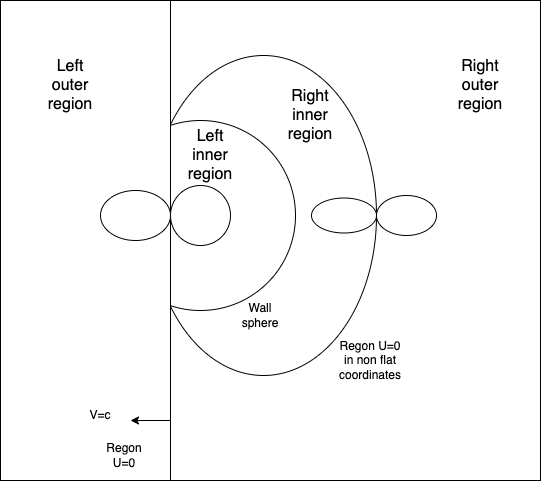}
    \footnotesize
   \caption{The four regions with their holes  $T>0$, adapted to flat coordinates in the inner left region. As these coordinates only make flatness explicit in that region, the other regions seems to be deformed
   and the spheres and circles also are not symmetric. That is the reason why the are drawn deformed. In addition, the larger right curve represents also a region $U=0$, it will be a line in other coordinates but not in the ones adapted
   to the left region.}
\end{figure}

\subsection{A more careful analysis of the resulting regions}

In all the four  regions defined  above, the coordinates (\ref{transfw})-(\ref{transfw4}) were chosen  in such a way that $T\to \pm\infty$ when $t\to \pm \infty$. The resulting coordinates
are continuous at all the borders of the regions. In all those regions $U<0$ and the horizon $U=0$ is a wall moving at light speed, running away from the points of the corresponding region.
A particle inside any region will never hit the wall at finite time $T$, unless a tachyon is involved in the game. This suggest, at least at first sight, that the outer regions and the inner regions are disconnected, as nothing cross the horizon. Instead, the two inner regions can communicate themselves, as the domain wall sphere reaches light speed only at asymptotic times. 

The previous conclusion is based on a choice of coordinates which is guided by the time direction defined by $t$. On the other hand, one may be skeptic that this time direction
is the true one, as the proper time is $T$, not $t$. In other words, a travel backwards in $t$ may not imply a backward travel in $T$ and vice versa. Therefore, one may explore the possibility of switching the signs of $U$ and $V$ in different regions, as this does not change the metric. However, this procedure does not help.
For instance, perform the switch $U_o\to- U_o$ at the outer region. Then the points are at $U_o>0$ and the wall seems to be pouncing over the region. However, in order 
to leave the metric invariant the shift $V_o\to -V_o$ should also be performed.  The simultaneous change $U_o\to -U_o$ and $V_o\to -V_o$ implies that the time direction $2T_o=V_o-U_o$ is changed.
If this time direction is changed, now the wall is again running away instead of moving forward. Again, the previous situation is repeated and the horizon $U=0$ is not to be crossed.

It is interesting to note that as part of the horizon $\Xi=2/ \mu$, the origin and the points $U=0$ corresponds by (\ref{transfw})-(\ref{transfw4}), to points such that either $t\to \pm \infty$ or $y\to \pm \infty$ or $z\to \pm \infty$. This suggest that those points parameterize the asymptotic region. However, consider the sphere  $T_i^2= X_i^2+Y_i^2+Z_i^2$. This sphere is also part of the horizon and is expanding at light speed. On one hand, it is clear that there are a lot of trajectories that can intersect this sphere at finite proper time. On the other hand, this region corresponds to $\alpha_i=0$ with $\beta_i\neq 0$. From (\ref{tata3}), which we write
here by convenience 
\be\lb{previ}
\alpha_i=\frac{2}{\mu}e^{-\frac{\mu t}{2}}\bigg(1-\frac{\mu \Xi}{2}\bigg),\qquad \beta_i=-\frac{2}{\mu\bigg(1-\frac{\mu \Xi}{2}\bigg)}e^{-\frac{\mu t}{2}},
\ee
it is seen that these points with $\alpha_i=0$ with $\beta_i\neq 0$ corresponds to trajectories that asymptotically tend to
$$
\bigg(1-\frac{\mu \Xi}{2}\bigg)e^{\frac{\mu t}{2}}\to \delta,
$$
with $\delta>0$ a constant, when $t\to +\infty$ while approaching the horizon $ \Xi\to 2/\mu$. From this it is concluded that these trajectories, even though are considered at $t\to \infty$, last
only a finite amount of proper time. This, fact suggests that the definition of the region of infinite is to be considered with care in this geometry, as an observer can cover all the positive times $t$ in few seconds when approaching the horizon by following those trajectories. It will be  seen in the next section that these trajectories are in fact light ones.


The next task is to discuss the geometry beyond the expanding sphere. In the following, two possibilities will be considered.

The first possibility can be described as follows. The region outside the hole corresponds to coordinates $\alpha_i >0$ and $\beta_i<0$. As the region in consideration corresponds to $U_i< 0$ the condition  $\beta_i<0$
is to be mantained.  It may be reasonable to postulate that the region inside the hole is such that $\alpha_i<0$ and $\beta_i<0$.  An inspection of (\ref{previ}) shows that if  the region $0<\mu \Xi<2$ is extended
 to points with $\mu \Xi>2$ and if the sign of $\beta_i$ is switched, then the values $\alpha_i<0$ and $\beta_i<0$ are obtained. By repeating the procedure
 done in (\ref{cf})-(\ref{flat}) adapted to the case $\beta\to -\beta$ leads to the following hole metric
\be\lb{cfdd}
g_h=\frac{1}{\beta^2}[-d\alpha d\beta+dy^2+dz^2].
\ee
The transformation 
\be\lb{transf2dd}
 \alpha=-V-\frac{Y^2+Z^2}{U},\qquad \beta=-\frac{1}{U}, \qquad y=\frac{Y}{U}, \qquad z=\frac{Z}{U},
\ee
with inverse
\be\lb{transfdd}
U=-\frac{1}{\beta}, \qquad V=-\alpha+\frac{y^2+z^2}{\beta}, \qquad Y=-\frac{y}{\beta}, \qquad Z=-\frac{z}{\beta},
\ee
converts the last metric into 
\be\lb{flatdd}
g_h=dU dV+dY^2+dZ^2,\qquad U=X-T, \qquad V=X+T.
\ee
Thus the metric is flat in this region. Note that the coordinates (\ref{transfw}) and (\ref{transfdd}) are different due to the sign of $\beta$. This means that an observer outside the hole that enters in the inner region
has to change a shuttle for seeing the flat metric. Explicitly the coordinates are 
$$
U_i=-\frac{\mu}{2}\bigg(1-\frac{\mu \Xi}{2}\bigg)e^{\frac{\mu t}{2} }, \qquad V_i=-\frac{2}{\mu}\bigg(1-\frac{\mu \Xi}{2}\bigg)e^{-\frac{\mu t}{2} }+\frac{\mu(y^2+z^2)}{2}\bigg(1-\frac{\mu \Xi}{2}\bigg)e^{\frac{\mu t}{2} }, 
$$
\be\lb{transfw11}
Y_i=-\frac{\mu y}{2}\bigg(1-\frac{\mu \Xi}{2}\bigg)e^{\frac{\mu t}{2} }, \qquad Z_i=-\frac{\mu z}{2}\bigg(1-\frac{\mu \Xi}{2}\bigg)e^{\frac{\mu t}{2} }.
\ee
These coordinates have the same functional form that  (\ref{transfw}).
However, since the region is naively the extension of $\mu \Xi<2$ to $2<\mu \Xi$ the discussion below (\ref{transfw}) shows that the coordinates do not respect the time orientation with respect to $t$.
A shift $U\to -U$ and $V\to -V$ can be made which respects time orientation. With the further change $Y\to -Y$ and $Z\to -Z$, a simple inspection show that these coordinates become the same as (\ref{transfw3}), since to change the sign is equivalent  to the change $1-\mu \Xi/2\to \mu \Xi/2-1$. The resulting coordinates share time orientation defined by $t$ namely, $T\to \infty$ when $t\to \infty$ and vice versa.
Furthermore these coordinates are continuous when crossing 
from the outer region with a trajectory 
$$
\bigg(1-\frac{\mu \Xi}{2}\bigg)e^{\frac{\mu t}{2}}=c,\qquad t\to \infty, \qquad \mu \Xi\to 2^-,
$$
and from the inner region with
$$
\bigg(\frac{\mu \Xi}{2}-1\bigg)e^{\frac{\mu t}{2}}=c,\qquad t\to \infty,\qquad \mu \Xi\to 2^+.
$$
Clearly the coordinates (\ref{transfw11}) takes the same  set of values of (\ref{transfw3}) for  $2<\mu \Xi$. This suggest that a particle which cross the hole does not enter to a compact region, 
instead they enter into a replica of the outer region. One may identify both regions as the same or instead to interpret this region as a true replica.  In either case, it is not difficult to see that the particles
become trapped at the hole location. The reason is that a particle which impacts the hole enters into a replica of the outer region. In this outer region, there is also a hole and furthermore the particle emerges
precisely from the boundary of this hole. However, since the hole moves at the light speed, the particle needs a speed larger than light to cross it. Therefore only a tachyon can cross it. The same follows in the other direction of movement, that is, a particle
entering from this replica into the inner region. Thus, the particle remains at the hole boundary forever. The same type of arguments follow for the other holes in the remaining regions.

Note that the particles trapped at the boundary of the hole, if they have non zero mass, move at the speed of light and have therefore infinite energy $E=m/\sqrt{1-v^2}$. Only massless particles in the region contribute with finite energy.
The gravitational field of such infinite energy configuration can be very chaotic and one may think that it will destroy easily the domain wall gravitational field. However, the hole moves at the speed of light in both regions and they never reach the wall.
The gravitational field arguably propagates at the same speed, thus it is also trapped at the hole boundary. It seems, at least classically that an enormous gravitational field appears that is in fact harmless, as nobody outside the hole can detect it.
This possibility assumes that the gravitational field  moves here at the light speed, although there is a discussion in the literature
that raises doubts about this fact in general curved space times \cite{arkani}-\cite{otras4}. This picture is of course classical, and the authors at the moment have no clear idea how quantum effects are included in this picture.

 The second possibility is to fill the holes with a flat space.
The interpretation is simply that the cartesian coordinates are not covering still the full geometry, and that the absence of matter inside the hole does not lead to any curvature. In fact, the reader can notice that one may extend the geometry and to assume that inside the hole may be anything, for instance and expanding curved universe. The true content of these holes is of course not completely determined by the present discussion. We suggest however that the first possibility  is a good candidate. 
 
\section{Geodesics and scalar fields on the geometry}

\subsection{Inner geodesics}\label{section:inner geodesics}
At first sight, the fact that the geometry is divided in flat regions suggest that the geodesics may be trivial. Before to enter into this subject, it is convenient to characterize the curves
joining two points $p$ and $q$ inside a circle. These curves will be assumed in the following to be composed by two straight lines. The first starts at $p$ and intersect the circle at some point $r$, and the second
goes from $r$ to $q$. The question is which of those possible curves minimizes the proper time which, if the velocity is constant, implies the minimization of the length of the curve.
This puts a condition on the intersection point $r$.
The first curve is
\be\lb{curo}
X=X_0+v(T-T_0)\cos \alpha, \qquad Y=Y_0+v (T-T_0)\sin\alpha.
\ee
This line forms an angle $\beta$ with the normal $\hat{n}$ to the sphere at the intersection point $r$. The value of $\beta$ is complicated of course, and depends on the direction $\alpha$.
Now draw the curve going backwards  forming the same angle $\beta$ with this normal $\hat{n}$, but reflected with respect of this normal. A simple exercise of drawing
and trigonometry shows that this curve is 
$$
\widetilde{X}=X_i-v(\widetilde{T}-T_i)\sin \gamma, \qquad \widetilde{Y}=Y_i-v (\widetilde{T}-T_i)\cos\gamma,\qquad \gamma=2\beta+\alpha.
$$
If $\beta$ is chosen such that this curve intersects $q$, then the resulting curve is the one that minimizes the elapsed time for traveling from $p$ to $q$. The angles that both the incident and reflected curve form with the normal of the circle at the intersection point are equal. 

The situation changes drastically if the circle moves. In this case, if the circle expands at high speed, the particle  may decide to intersect it as fast as possible and to bounce back
in the direction to the point $q$. Otherwise the circle will be very large and long  distance is to be traveled, which makes the proper time of the trajectory very large. Thus, it is not warranted that $\beta$ is conserved for a moving circle or sphere. In fact, this can be  calculated explicitly. The intersection time $T_1$ of the trajectory (\ref{curo}) with the domain wall sphere $1+T_1^2= X_1^2+Y_1^2+Z_1^2$ is given by
$$
T_1-T_0=\frac{v(T_0-X_0\cos\alpha-Y_0 \sin\alpha)
\pm \sqrt{v^2(T_0-X_0\cos\alpha-Y_0 \sin\alpha)^2-(1-v^2)(1-X_0^2-Y_0^2+2 T_0^2)}}{1-v^2}.
$$
By use of (\ref{curo}) the intersection position $X_1$ and $Y_1$ is also found. The proper time elapsed to reach the final position $X_f$ and $Y_f$ is given by
$$
\Delta \tau=\sqrt{1-v^2}(T_1-T_0)+\sqrt{1-v^2}\frac{\sqrt{(X_f-X_1)^2+(Y_f-Y_1)^2}}{v}.
$$
The critical point is found by imposing that the derivative with respect to $\alpha$ of the last expression is zero namely
$$
\frac{dT_1}{d\alpha}\bigg[1-\frac{\cos\alpha (X_f-X_1)+\sin\alpha (Y_f-Y_1)}{\sqrt{(X_f-X_1)^2+(Y_f-Y_1)^2}}\bigg]=0.
$$
There are several possible solutions. A first possibility is that $T_1$ is minimal, which implies that the particle tries to cross the wall as fast as possible. It is a simple exercise to check that in 
a generic case, there are more than one solution of this condition. The other solution
is that the term in brackets is zero. One of those solutions is the global minimum, and this has to be checked separately for every given initial condition. However, depending on these conditions, two  or more local minima may correspond to curves
that elapse the same proper time.
In this case, the initial and final points $p$ and $q$ are conjugate to each other. 

For light trajectories, the trajectory between two points is the one that makes the distance minimal. We leave for the reader to work out this situation. In addition, note that the trajectories just described may never reach the wall or may fall
into the expanding small hole present in the geometry, discussed in the previous section.

Described in the Cartesian coordinates as above, this proper time minimizing curve seems to be have discontinuous derivatives with respect to the time parameter $T$ at the border of the sphere.
This is not necessarily in discrepancy  with the fact that a geodesic is a $C^2$ curve without conjugate points. This famous result follows from studying geodesic deviation, which involves
the curvature, and its applicability may be doubtful at points where the curvature is singular, such as the position of the wall \cite{texto1}-\cite{texto2}. On the other hand, the discontinuity may be an artifact of the choice
of the coordinates. It may be the case that there is a coordinate system for which the curve looks smooth. Thus it may be interesting to see if this coordinate system exists, a candidate may be the original one $(t, x, y, z)$ defined in the metric (\ref{domwall}). This incomplete metric is enough for describing this region, as the outer region is disconnected from the inner one.

If $z$ is assumed to be constant then the geodesic is given by a curve $(t(s), x(s), y(s))$ such that
\be\lb{domwall2}
-e^{-\mu |x|} \bigg(\frac{dt}{ds}\bigg)^2+e^{-\mu |x|}\bigg(\frac{dx}{ds}\bigg)^2+e^{\mu(t-|x|)}\bigg(\frac{dy}{ds}\bigg)^2=\epsilon.
\ee

For null geodesics $\epsilon=0$ while for time like geodesics $s=\tau$ with $\tau$ the proper time, and $\epsilon=1$. 
Consider the region $x>0$. The geodesic equation for $x(s)$ becomes
$$
\frac{d^2x}{ds^2}-\frac{\mu}{2}\bigg(\frac{dx}{ds}\bigg)^2-\frac{\mu}{2}\bigg(\frac{dt}{ds}\bigg)^2+\frac{\mu e^{\mu t}}{2}\bigg(\frac{dy}{ds}\bigg)^2=0.
$$
This can be rewritten by use of (\ref{domwall2}) as
\be\lb{geo1}
\frac{d^2x}{ds^2}-\mu\bigg(\frac{dx}{ds}\bigg)^2=-\frac{\epsilon\mu}{2} e^{\mu x},
\ee
giving decoupled equations for $x(s)$.

An universal form for this equation that is valid on both sides is 
\be\lb{geo3}
\frac{d^2x}{ds^2}-\mu\; \text{sign}(x)\bigg(\frac{dx}{ds}\bigg)^2=-\frac{\epsilon\mu}{2} e^{\mu |x|}.
\ee

From (\ref{geo3}) it can be seen that continuous solutions $x(s)$ must also have continuous first derivative everywhere, and in particular at the wall domain, $x=0$. The discontinuity of $\text{sign}(x)$ will not affect the continuity properties of $x'(s)$, this can be seen from the equation by doing a measure zero integration around the wall domain.

Consider first null geodesics. In this case the equation is 
$$
\frac{d}{dx}\bigg(\frac{dx}{ds}\bigg)-\mu\; \text{sign}(x) \frac{dx}{ds}=0.
$$
The integration of the last equation, if the initial point and derivative are $x_0>0$ and $x'_{0}<0$ is given by\footnote{In this case light travels from the right side of the wall towards it, and, as we'll see next, it pass through. Initials conditions where $x_0<0$ and $x'_{0}>0$ leads to similar solutions from left to right of the wall domain.}:
\begin{align}	
x_{+}(s)&=x_{0}-\frac{1}{\mu}\log \left| 1-\mu x'_0 s \right|=-\frac{1}{\mu} \log \left| e^{- x_0 \mu } \left( 1-\mu x'_0 s \right) \right|,  &\quad &0<s<s_c  &\quad (x>0), \nonumber\\
\label{integra}\\
x_{-}(s)&=\frac{1}{\mu } \log \left| 1+ \mu x'_c (s-s_{c}) \right|=\frac{1}{\mu }{\log \left| 2- e^{- x_0 \mu } \left(1- \mu  x'_0 s \right)\right| } , &\quad &s_c<s<s_{\infty}  &\quad (x<0),\nonumber
\end{align}
where $s_c=(1-e^{\mu  x_{0} })/{\mu  x'_0}$ is the value of the parameter where the wall domain is reached ($x_{+}(s_{c})=x_{-}(s_{c})=0$), $s_{\infty} = {\left(1-2e^{\mu  x_0}\right)}/{\mu  x'_0}$ the value where $\left|x\right| \rightarrow \infty$, and $x'_c =x'_{+} (s_c)= x'_0 e^{- \mu  x_0}\text{ }(=x'_{-} (s_c))$.

Notice that at $s_\infty$ is where it would be the horizon in the $\Xi$ coordinate, and beyond $s_\infty$ it would be as entering the extended outer region of the space defined by $\Xi$.

To completely solve for null like geodesics, it can be seen that the universal form of the rest of the geodesics equations at both sides of the wall are\footnote{Without loss of generality we give the solutions at the $z=0$ plane.}
\begin{align}	
	&\frac{d^2t}{ds^2}-\mu\frac{dt}{ds}\frac{dx}{ds}+\frac{\mu}{2} \bigg(\frac{dt}{ds}\bigg)^2-\frac{\mu}{2} \text{sign}(x) \bigg(\frac{dx}{ds}\bigg)^2=0, \label{eqn:ts}\\
	\nonumber\\
	&\frac{d^2y}{ds^2}+\mu\frac{dy}{ds} \left(\frac{dt}{ds}-\text{sign}(x)\frac{dx}{ds}\right)=0.\label{eqn:ys}
\end{align}

Here also, for continuous $t(s)$ and $y(s)$ the equations tell us that they must have continuous first derivatives with respect to $s$.

With $x(s)$ already known, (\ref{eqn:ts}) can be solved changing to the variable ${t'(s) - \text{sign}(x) x'(s)}$ to give
\begin{align}	
	t_{+}(s) &=  x_{+}(s) + t_{0}-x_{0} + \frac{2}{\mu} \log \left| \frac{\mu}{2} \left( t'_{0}-x'_{0} \right) s +1\right| &\quad &0<s<s_c \nonumber\\
	&=t_0 + \frac{1}{\mu} \ln \left|\frac{\left(\frac{\mu}{2} \left( t'_{0}-x'_{0} \right) s+1\right)^2}{1-\mu x_0^{\prime} s}\right|, \label{eqn:solts+}\\
	\nonumber\\
	t_{-}(s) &=  -x_{-}(s) + t_{c} + \frac{2}{\mu} \log \left| \frac{\mu}{2} \left( t'_{c}+x'_{c} \right) \left(s - s_{c}\right) +1\right| &\quad &s_c<s<s_{\infty}\nonumber\\
	&=t_c + \frac{1}{\mu} \ln \left|\frac{\left(\frac{\mu}{2} \left( t'_{c}+x'_{c} \right) \left( s-s_{c} \right)+1\right)^2}{1+\mu x_c^{\prime} \left( s-s_{c} \right)}\right|, \label{eqn:solts-}
\end{align}
where\footnote{Notice that here the sub-indexes $(+)$ and $(-)$ in their respective $s$ region, are inherited by the positivity and negativity of $x(s)$ but doesn't say anything about the sign of $t(s)$ itself.}, $t_c=t_{+} (s_c)\text{ }(=t_{-} (s_c))$ and $t'_c=t'_{+} (s_c)\text{ }(=t'_{-} (s_c))$. Also $\left|t\right| \rightarrow \infty$ as $s \rightarrow s_{\infty}$, and it is clear that light can't cross the horizon $\left| \Xi \right|=2/\mu$.

Now with the known solutions of $x(s)$ and $t(s)$, equation (\ref{eqn:ys}) can be solved to find
\begin{align}	
	y_{+}(s)&=y_0 - \frac{2y_0^{\prime}}{\mu \left( t'_{0}-x'_{0} \right)}\left[\frac{1}{\frac{\mu}{2} \left( t'_{0}-x'_{0} \right) s+1}-1\right], \label{eqn:solys+}\\
	\nonumber\\
	y_{-}(s)&=y_c - \frac{2y_c^{\prime}}{\mu \left( t'_{c}+x'_{c} \right)}\left[\frac{1}{\frac{\mu}{2} \left( t'_{c}+x'_{c} \right) \left( s-s_{c} \right)+1}-1\right],\label{eqn:solys-}
\end{align}
where $y_c=y_{+} (s_c)\text{ }(=y_{-} (s_c))$ and $y'_c=y'_{+} (s_c)\text{ }(=y'_{-} (s_c))$.

In the case of time like geodesics $\epsilon=1$, the resulting equation (\ref{geo3}) is more complex, and it is known as a Riccati type equation. Again, continuity of the function and its derivatives can be imposed. In order to find the solution of (\ref{geo1}) it is convenient to define
$$
G(s)= e^{\mu x},\qquad x=\frac{1}{\mu} \log G.
$$
leading to 
$$
\frac{d^2G}{ds^2}=-\frac{\epsilon \mu}{2} G^2.
$$
The conserved energy is 
$$
\bigg(\frac{dG}{ds}\bigg)^2+\frac{\epsilon \mu}{3}G^3=c.
$$
The integration leads to an expression in terms of hypergeometric functions
$$
s-d=_2F_1(\frac{1}{3}, \frac{1}{2}, \frac{4}{3}, \frac{\mu \epsilon e^{3\mu x}}{c}) e^{\mu x}.
$$
This is an implicit and complicated expression defining $x(s)$, with two integration constants $c$ and $d$ if $\epsilon=1$.
The coordinates still are continuous in this case.

For $t(s)$ the equation is 
$$
\frac{d^2t}{ds^2}-\mu\frac{dt}{ds}\frac{dx}{ds}+\frac{\mu e^{\mu t}}{2} \bigg(\frac{dy}{ds}\bigg)^2=0.
$$
As $x(s)$ is continuous it is clear that $t(s)$ can also be required to be continuous.
By taking into account (\ref{domwall2}) leads to
$$
\frac{d^2t}{ds^2}-\mu\frac{dt}{ds}\frac{dx}{ds}+\frac{\mu}{2} \bigg(\frac{dt}{ds}\bigg)^2-\frac{\mu}{2} \bigg(\frac{dx}{ds}\bigg)^2=-\frac{\epsilon\mu}{2} e^{\mu x},
$$
or alternatively as
$$
\frac{d}{ds}\bigg(\frac{dt}{ds}-\frac{dx}{ds}\bigg)+\frac{\mu}{2} \bigg(\frac{dt}{ds}-\frac{dx}{ds}\bigg)^2+\frac{d^2x}{ds^2}-\mu \bigg(\frac{dx}{ds}\bigg)^2=-\frac{\epsilon\mu}{2} e^{\mu x},
$$
The use of (\ref{geo1}) simplifies the last expression to
\be\lb{geo2}
\frac{d}{ds}\bigg(\frac{dt}{ds}-\frac{dx}{ds}\bigg)+\frac{\mu}{2} \bigg(\frac{dt}{ds}-\frac{dx}{ds}\bigg)^2=0,
\ee
Solving for $t'(s)-x'(s)$ and then integrating, the solution of the last equation is:
\be\lb{wrong}
t-x=-\frac{2}{\mu}\log(1+\frac{\mu b s}{2})+b.
\ee
Similar considerations can be imposed on the right region, but the corresponding quantity is $t+x$. The integration constant $b$ in the last expression is not necessarily the same as the $a$
in (\ref{integra}). In general, given some particular initial conditions it may be seen from these formulas that  $x\to \pm \infty$ for finite values of $s$.
However (\ref{wrong}) also implies that $t\to \infty$ at this finite parameter value. At first sight this does not mean that a particle can not cross the horizon, although
it suggests the possibility. The results of the previous sections are those which shows that this possibility is realized.

The remaining function $y(s)$ can be found from (\ref{domwall2}), but this integration is rather complicated. Still, it is continuous and with continuous
derivatives when crossing the wall. 

The previous discussion shows that there exist curves satisfying the geodesic equation on both sides which are continuous and with continuous derivatives at the wall, when described
with an specific coordinate system $x$, $y$, $z$ and $t$. When going to the Cartesian coordinates, they become continuous but with discontinuous first derivatives. This can be easily seen by expressing (\ref{transfw}) in these coordinates as:
$$
U_i=-\frac{\mu}{2}e^{\frac{\mu (t-| x | )}{2} }, \qquad V_i=-\frac{2}{\mu}e^{-\frac{\mu (t+| x| )}{2} }+\frac{\mu(y^2+z^2)}{2}e^{\frac{\mu (t-|x| )}{2} }, $$
\be\lb{transfo}
Y_i=-\frac{\mu y}{2}e^{\frac{\mu (t-| x| )}{2} }, \qquad Z_i=-\frac{\mu z}{2}e^{\frac{\mu (t-| x| )}{2} }.
\ee

As $t$ and $x$ are continuous as functions of the parameter $s$, the geodesics in these coordinates are also continuous, but, regardless that $t'(s)$ and $x'(s)$ are continuous, now the Cartesian coordinates have discontinuous first derivatives with respect to $s$ at the wall, as $\frac{d}{ds}(t-| x | )$ is discontinuous there.

Even thought these trajectories are $C^2$ in coordinates $(t(s), x(s), y(s), z(s))$, it is not smooth at this intersection region when written in Cartesian coordinates. They present a Snell like deviation, but the angle may depend on the initial conditions and time.
The same follows for the proper time minimizing curve described at the beginning of this section.
Still, this does not prove that both curves  are the same. We have no analytic proof of that, as we were unable to integrate $t(x)$, $x(s)$ and $y(s)$ in complete explicit form.
In any case, we strongly suggest to the reader to employ the proper time minimizing curve described at the beginning of this subsection in cartesian coordinates in any concrete task.

\subsection{Scalar fields on the space time}

The characterization given above of the asymptotic region of the domain wall geometry is important
for studying the dynamics of quantum fields on the geometry. In the present section some comments will be made.
A deeper analysis of those matters will be the topic of a separate publication.

Given the fact that the regions of the space time are all locally flat, one may consider a monochromatic wave filing the right inner region
\be\lb{1}
\phi_I= e^{ik_-U+i k_+ V+i k_y Y+ik_z Z}.
\ee
In coordinates $x$, $y$, $z$ and $t$ this reads
\be\lb{I}
\phi_i= e^{ik_-\frac{\mu}{2}e^{-\frac{\mu}{2} (t+x)}+i k_+ \bigg[\frac{2}{\mu}e^{\frac{\mu}{2} (t-x)}-\frac{\mu(y^2+z^2)}{2}e^{-\frac{\mu}{2} (t+x)}\bigg]-i k_y \frac{\mu y}{2}e^{-\frac{\mu}{2} (t+x)}-ik_z \frac{\mu z}{2}e^{-\frac{\mu}{2} (t+x)}}.
\ee
On the left hand side one may postulate that the solution is
\be\lb{2}
\widetilde{\phi}_i= e^{ik_-\frac{\mu}{2}e^{-\frac{\mu}{2} (t-x)}+i k_+ \bigg[\frac{2}{\mu}e^{\frac{\mu}{2} (t+x)}-\frac{\mu(y^2+z^2)}{2}e^{-\frac{\mu}{2} (t-x)}\bigg]-i k_y \frac{\mu y}{2}e^{-\frac{\mu}{2} (t-x)}-ik_z \frac{\mu z}{2}e^{-\frac{\mu}{2} (t-x)}}.
\ee
However, we believe that this result is not the correct answer yet. These solutions are continuous with discontinuous derivatives in $x$ along the wall and along the horizon singularities. Note that there exist an analysis of the scalar and Dirac fields
in this geometry performed in \cite{bezerra}, where the eigenfunctions  in the inner geometry were found explicitly. The results of that reference suggest this discontinuous behavior
along the horizon. In addition, our intuition is that the formulas (\ref{1})-(\ref{2}) are not the end of the story. The point is that the eikonal limit should be described by the geodesic direction, and this direction was shown in the previous section to change when crossing the wall. Thus, we suspect that the direction defined by $k_i$ may change when crossing the wall.
There may be a reflected wave as well, and both the reflected and transmitted wave may be a complicated Fourier superposition. The continuity condition should hold, and 
the action corresponding to the full field configuration should be minimal. These two conditions may be enough for determining this Fourier expansion, but it will be considered separately,
as the task may be not as simple as looks like. 

In fact, the complication when dealing with scalar fields can be stated simple. When the boundary conditions are not static, even the Schrodinger equation for a free particle in a moving infinite potential well can be non trivial. Thus, the solution of the free Klein Gordon equation with such moving  boundary is to be studied with care.
 
A further comment is that the Green function corresponding to a source on the left region I
is given by
$$
G(T', X'_i, T_1, X_i )=i \theta(T'-T_1)\int \frac{d^3k}{2(2\pi)^3\omega_k} e^{-i\omega_k (T'-T_1)} e^{i k_x(X'-X_1)}
e^{i k_y(Y'-Y)}e^{i k_z(Z'-Z)}
$$
$$
+i \theta(T_1-T')\int \frac{d^3k}{2(2\pi)^3\omega_k} e^{i\omega_k (T'-T_1)} e^{i k_x(X'-X_1)}
e^{i k_y(Y'-Y)}e^{i k_z(Z'-Z)}.
$$
Here one may replace the coordinates $U$, $V$, $Y$ and $Z$ in terms of $x$, $y$, $z$ and $t$.
By simply changing $x$ to $-x$ in the left hand side, the monochromatic waves on every singularity region would be
$$
G(T', X'_i, T_2, X_i )=i \theta(T'-T_2)\int \frac{d^3k}{2(2\pi)^3\omega_k} e^{-i\omega_k (T'-T_2)} e^{i k_x(X'-X_2)}
e^{i k_y(Y'-Y)}e^{i k_z(Z'-Z)}
$$
$$
+i \theta(T_2-T')\int \frac{d^3k}{2(2\pi)^3\omega_k} e^{i\omega_k (T'-T_2)} e^{i k_x(X'-X_2)}
e^{i k_y(Y'-Y)}e^{i k_z(Z'-Z)}.
$$
However, this naive formula does not work, as it violates causality. The point is that this formula represents an image source at a point with the 
same coordinates as the original, but in the other side. This expression should vanish for space like separated points. However, as indicated in the text the coordinates
on the left and the right ranges the same set of values and thus the last expression does not describe well the true separation between the points. This separation depends
on the distance to the wall $X^2+Y^2+Z^2=1+T^2$ and the last formula does not reflect this distance. It will be interesting
to find the true expression for the Green functions in different regions, since this may be particularly important for calculating the vacuum polarization in this geometry and its analytic properties \cite{caus2}-\cite{hollowood}. This will be considered elsewhere, but it surely will involve the geodesic distance if the path integral methods of those references are to be employed.

\section{Discussion}

The present letter, by use of Taub coordinates, suggests that given a domain wall with energy momentum tensor $T_{\mu\nu}=\mu (1,0,1,1)$ in GR, the resulting geometry possess two horizons, one at the left
and one on the right of the wall that can not be crossed from inside or outside. This feature is different than for black holes, where the horizon can be crossed at finite proper time. In addition, there appear at some time of the evolution
four holes growing at the speed of light, which these coordinates do not cover. It was suggested here that there are coordinates inside this hole 
that convert the metric into the Minkowski flat line element, but which are different than the outside ones. Furthermore, when the particles collide with one of such holes, they are  adsorbed at the boundary.

The boundary of the holes described above corresponds to travels which, in the coordinates of references \cite{vilenkin1}-\cite{ipser}, reach asymptotic values which are in fact at finite proper time from any point of the region.

 Furthermore, it was shown along the text that the internal regions on the left and the right of the wall communicate
each other, since the separating domain wall runaways with speed smaller than the light from the point inside the corresponding region. 

In addition, even though that in patches the geodesics are simply straight lines, there appears an Snell type of deviation at the wall sphere for classical particles, and the corresponding angle variation is time and initial conditions dependent.
This dependence is what breaks Poincare invariance on the otherwise flat geometry.

Another interesting point mentioned in the text is that, given a scalar charge on the left side, the particles on the right feels the corresponding field as if there were an image source on the right region. This in particular may lead to a non trivial self force for a given field, indicating that the geometry is not flat globally. However, the previous formulas do not respect causality yet, and have to be further worked out. The problem is that in the eikonal limit the geodesics dictate the direction of propagation, and as stated
in the previous paragraph, these geodesics are non trivial.

The scattering problems for fields with different spin in this geometry may be non trivial as well, even though the regions are flat in patches, as the boundary is not so trivial to be described. In particular, the points that are apparently asymptotic but are at finite proper time distance of the points of the space time, which were described along the text, may introduce additional complications.

Furthermore, as the presence of the wall breaks traslational symmetry as well as time shifts invariance, it may be interesting to study the refraction index 
$n(x^\mu, \omega)$ for light propagation along the domain wall and to understand how this breaking influences on the otherwise analytic properties of this index as a function of the complex frequency values  \cite{caus2}-\cite{hollowood}.
These matters will be a topic of a separate publication.

\section*{Acknowledgments}
Both authors are supported by CONICET, Argentina and O. P. S is also supported by the Grant PICT 2020-02181.

\end{document}